\documentclass[11pt,a4paper]{article}
\usepackage[hyperref]{naaclhlt}
\usepackage{times}
\usepackage{latexsym}
\usepackage{url}

\usepackage{tikz}
\usetikzlibrary{arrows,positioning}
\usepackage{amsmath}
\usepackage{amsthm}
\usepackage{amssymb}
\usepackage{multirow}
\usepackage{color,soul}
\usepackage{paralist}
\usepackage{booktabs}
\usepackage{xcolor}
\usepackage{units}
\usepackage{bbding}
\usepackage{verbatim}

\aclfinalcopy

\theoremstyle{plain}
\newtheorem{defn}{Definition}

\newtheorem{thm}{Theorem}
\newtheorem{corollary}{Corollary}
\newtheorem{conjecture}{Conjecture}
\theoremstyle{definition}
\newtheorem{example}{Example}

\title{The problem with probabilistic DAG automata for semantic graphs}

\author{Ieva Vasiljeva\Thanks{~Equal contribution. Work while Ieva Vasiljeva was at the University of Edinburgh} \and Sorcha Gilroy$^*$ \and Adam Lopez\\
Institute for Language, Cognition, and Computation\\
School of Informatics\\
University of Edinburgh\\ 
}

\date{}
\begin{document}
\maketitle
\begin{abstract}
Semantic representations in the form of directed acyclic graphs (DAGs) have been introduced in recent years, and to model them, we need probabilistic models of DAGs. One model that has attracted some attention is the DAG automaton, but it has not been studied as a probabilistic model. We show that some DAG automata cannot be made into useful probabilistic models by the nearly universal strategy of assigning weights to transitions. The problem affects single-rooted, multi-rooted, and unbounded-degree variants of DAG automata, and appears to be pervasive. It does not affect planar variants, but these are problematic for other reasons. 
\end{abstract}

\section{Introduction}
Abstract Meaning Representation (AMR; \citealt{amr}) has prompted a flurry of interest in probabilistic models for semantic parsing. AMR annotations are directed acyclic graphs (DAGs), but most probabilistic models view them as strings \citep[e.g.][]{vanNoord} or trees \citep[e.g.][]{Flanigan+etal:2016:naacl}, removing their ability to represent coreference---one of the very aspects of meaning that motivates AMR. Could we we instead use probabilistic models of DAGs?

To answer this question, we must define probability distributions over sets of DAGs. For inspiration, consider probability distributions over sets of strings or trees, which can be defined by weighted finite automata \citep[e.g.][]{Mohri+etal:2008:hspsc,May:2010:EIT:1858681.1858789}: a finite automaton generates a set of strings or trees---called a language---and if we assume that probabilities factor over its transitions, then any finite automaton can be weighted to define a probability distribution over this language. This assumption underlies powerful dynamic programming algorithms like the Viterbi, forward-backward, and inside-outside algorithms.

What is the equivalent of weighted finite automata for DAGs? There are several candidates \citep{chiang,drewes_rock,mol_17}, but one appealing contender is the \emph{DAG automaton} \citep{Quernheim:2012:TPA:2392936.2392948} which generalises finite tree automata to DAGs explicitly for modeling semantic graphs. These DAG automata generalise an older formalism called \textit{planar DAG automata} \citep{kamimura} by adding weights and removing the planarity constraint, and have attracted further study \citep{DBLP:conf/lata/BlumD16,drewes_mol}, in particular by \citet{CL}, who generalised classic dynamic programming algorithms to DAG automata. But while 
\citet{Quernheim:2012:TPA:2392936.2392948} clearly intend for their weights to define probabilities, they stop short of claiming that they do, instead ending their paper with an open problem: ``\emph{Investigate a reasonable probabilistic model.}''
 
We investigate probabilistic DAG automata and prove a surprising result: \textbf{For some DAG automata, it is impossible to assign weights that define non-trivial probability distributions}.
We exhibit a very simple DAG automaton that generates an infinite language of graphs, and for which the only valid probability distribution that can be defined by weighting transitions is one in which the support is a single DAG, with all other graphs receiving a probability of zero. 

Our proof relies on the fact that a non-planar DAG automaton generates DAGs so prolifically that their number grows factorially in their size, rather than exponentially as in other automata. It holds for DAG automata that allow multiple roots or nodes of unbounded degree. But it breaks down when applied to the planar DAGs of \citet{kamimura}, which are nevertheless too restrictive to model semantic graphs. Our result does not mean that it is impossible to define a probability distribution for the language that a DAG automaton generates. But it does mean that this distribution does not factor over the automaton's transitions, so crucial dynamic programming algorithms do not generalise to DAG automata that are expressive enough to model semantic graphs.

\section{DAGs, DAG automata, and probability\label{sec:prelim}}

We are interested in AMR graphs like the one below on the left for ``Rahul bakes his cake'', which
represents entities and events as nodes, and relationships between them as edges. Both nodes and edges have labels, representing the type of an entity, event, or relationship. But the graphs we model will only have labels on nodes. These node-labeled graphs can simulate edge labels using a node with one incoming and one outgoing edge, as in the graph on the right.

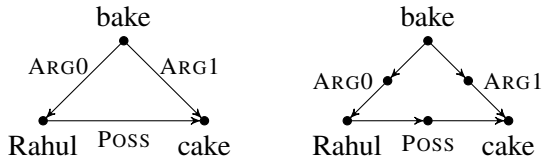
\begin{center}
\begin{tikzpicture}[>=stealth',auto,node distance=1.5cm,
  square node/.style={draw,line width=0.25mm,rectangle,inner sep=2pt}
,dot node/.style={draw,circle,minimum size=1.1mm,fill,inner sep=0pt,outer sep=0pt}]
\node[dot node,label=above:{bake}] (miss) {};
\node[dot node, below left of = miss,label=below:{Rahul}] (anna) {};
\node[dot node, below right of = miss,label=below:{cake}] (cat) {};

\path[every node/.style={font=\small}]

(miss) edge[->] node[right,xshift=-0.2cm,yshift=0.2cm] {\textsc{Arg1}} (cat)
edge[->] node[left,xshift=0.2cm,yshift=0.2cm] {\textsc{Arg0}} (anna)
(anna) edge[->] node[below] {\textsc{Poss}} (cat);

\node[right of=miss,node distance=4cm,dot node,label=above:{bake}] (miss) {};
\node[dot node, below left of = miss,label=below:{Rahul}] (anna) {};
\node[dot node, below right of = miss,label=below:{cake}] (cat) {};

\node[dot node, below left of = miss,node distance=7.5mm,label=left:{\small\textsc{Arg0}}] (arg1) {};
\node[dot node, below right of = miss,node distance=7.5mm,label=right:{\small\textsc{Arg1}}] (arg0) {};
\node[dot node, below of=miss,node distance=10.6mm,label=below:{\small\textsc{Poss}}] (poss) {};

\path[every node/.style={font=\small}]

(miss) edge[->] (arg0) edge[->] (arg1)
(arg1) edge[->] (anna)
(arg0) edge[->] (cat)
(anna) edge[->] (poss)
(poss) edge[->] (cat);

\end{tikzpicture}
\end{center}
%
%

\begin{defn}
A node-labeled directed \textbf{graph} over a label set $\Sigma$ is a tuple $G = (V,E,\text{lab},\text{src},\text{tar})$ where $V$ is a finite set of nodes, $E$ is a finite set of edges, lab$:V \rightarrow \Sigma$ is a function assigning labels to nodes, src$:E \rightarrow V$ is a function assigning a source node to every edge, and tar$:E \rightarrow V$ is a function assigning a target node to every edge.
\label{defn:graph}
\end{defn}

Sometimes we will discuss the set of edges coming into or going out of a node, so we define functions \textsc{In}$:V \rightarrow E^*$ and \textsc{Out}$:V \rightarrow E^*$.
\begin{align*}
\textsc{In}(v) &= \{e \mid \text{tar}(e) = v\}\\
\textsc{Out}(v) &= \{e \mid \text{src}(e) = v\}
\end{align*}
A node with no incoming edges is called a \textbf{root}, and a node with no outgoing edges is called a \textbf{leaf}.
The \textbf{degree} of a node is the number of edges connected to it, so the degree of $v$ is $|\textsc{In}(v)\cup \textsc{Out}(v)|$.

A \textbf{path} in a directed graph from node $v$ to node $v'$ is a sequence of edges $(e_1,\dots,e_n)$ where src$(e_1) = v$, tar$(e_n) = v'$ and src$(e_{i+1}) = $ tar$(e_{i})$ for all $i$ from $1$ to $n-1$. A \textbf{cycle} in a directed graph is any path in which the first and last nodes are the same (i.e., $v = v'$). A directed graph without any cycles is a \textbf{directed acyclic graph (DAG)}.

A DAG is \textbf{connected} if every pair of its nodes is connected by a sequence of edges, not necessarily directed. 
Because DAGs do not contain cycles, they must always have at least one root and one leaf, but they can have multiple roots and multiple leaves. However, our results apply in different ways to single-rooted and multi-rooted DAG languages, so, given a label set $\Sigma$, we distinguish between the set of all connected DAGs with a single root, $\mathcal{G}_\Sigma^1$; and those with one or more roots, $\mathcal{G}_\Sigma^*$.
\subsection{DAG automata}

Finite automata generate strings by transitioning from state to state. Top-down tree automata generalise string finite automata by transitioning from a state to an ordered sequence of states, generating trees top-down from root to leaves; while bottom-up tree automata transition from an ordered sequence of states to a single state, generating trees bottom-up from leaves to root. The planar DAG automata of \citet{kamimura} generalise tree automata, transitioning from one ordered sequence of states to another ordered sequence of states (Section \ref{sec:planar}). Finally, the DAG automata of \citet{Quernheim:2012:TPA:2392936.2392948} transition from \emph{multisets} of states to multisets of states, rather than from sequences to sequences, and this allows them to generate non-planar DAGs. We summarise the differences below.

\begin{center}\small
\begin{tabular}{ccc}
\toprule
Automaton & Transitions & Example \\
\midrule
string & one-to-one & $p \rightarrow p'$\\
top-down tree & one-to-many & $p \rightarrow (p',q')$\\
bottom-up tree & many-to-one & $(p',q')\rightarrow p$ \\
planar DAG & many-to-many & $(p,q) \rightarrow (p',q')$\\
non-planar DAG & many-to-many & $\{p,q\} \rightarrow \{p',q'\}$\\
\bottomrule
\end{tabular}
\end{center}

\begin{figure*}[t]
\begin{center}
\begin{tikzpicture}[>=stealth',auto,node distance=8mm,
  square node/.style={draw,line width=0.25mm,rectangle,inner sep=2pt}
,dot node/.style={draw,circle,minimum size=1.1mm,fill,inner sep=0pt,outer sep=0pt},
,derives/.style={->,double,lightgray}]

\node (q1') {};
\node[dot node, left of = q1',label=left:{$b$}] (b1') {};
\node[dot node, above of = b1',label=left:{$a$}] (a') {};
\node[dot node, below of = b1',label=left:{$b$}] (b2') {};
\node[dot node, below of = b2',label=left:{$c$}] (c') {};
\node[right of = b2'] (q2') {};
\node[above right of = c',yshift= -0.25] (p') {};

\coordinate[left of = b1',xshift = 3mm] (en_arr');
\coordinate[left of = en_arr'] (st_arr');
\node[left of = st_arr', node distance=1mm] (q1'') {};
\node[dot node, left of = q1'',label=left:{$b$}] (b1'') {};
\node[dot node, above of = b1'',label=left:{$a$}] (a'') {};
\node[below of = b1''] (p2'') {};

\coordinate[left of = b1'', xshift = 0.3cm] (en_arr'');
\coordinate[left of = en_arr'', xshift = 0.3cm] (st_arr'');
\node[left of = st_arr'',node distance=2mm] (p1''') {};
\node[dot node, above of = p1''',label=left:{$a$}] (a''') {};

\coordinate[left of=st_arr'',xshift=3mm] (en_arr''');
\coordinate[left of=en_arr''',xshift=3mm] (st_arr''');

\node[right of = c'] (lab_c) {(\romannumeral 3)};
\node[below of = a'',node distance=24mm] (lab_b) {(\romannumeral 2)};
\node[below of = a''',node distance=24mm] (lab_a) {(\romannumeral 1)};

\path[every node/.style={font=\sffamily\small}]

(a') edge[->] (b1')
(b1') edge[->] (b2')
(b1') edge[->] node[above] {\textcolor{red}{$q$}} (q1')
(b2') edge[->] (c')
(b2') edge[->] node[above] {\textcolor{red}{$q$}} (q2')
(c') edge[->] node[right,xshift=1.5mm,yshift=2mm] {\textcolor{red}{$p'$}} (p')

(st_arr') edge[derives] node[below] {$t_2, t_3$}(en_arr')

(a'') edge[->] (b1'')
(b1'') edge[->] node[above] {\textcolor{red}{$q$}} (q1'')
(b1'') edge[->] node[right] {\textcolor{red}{$p$}} (p2'')

(st_arr'') edge[derives] node[below] {$t_2$} (en_arr'')
(st_arr''') edge[derives] node[below] {$t_1$} (en_arr''')

(a''') edge[->] node[right] {\textcolor{red}{$p$}} (p1''');

\coordinate[right of = q1',node distance=1mm] (st_arr);
\coordinate[right of = st_arr,node distance=5mm] (en_arr);
\path[every node/.style={font=\sffamily\small}]
(st_arr) edge[derives] node[below] {$t_4$}(en_arr);

\node[right of = en_arr,node distance=14mm] (q1') {};
\node[dot node, left of = q1',label=left:{$b$}] (b1') {};
\node[dot node, above of = b1',label=left:{$a$}] (a') {};
\node[dot node, below of = b1',label=left:{$b$}] (b2') {};
\node[dot node, below of = b2',label=left:{$c$}] (c') {};
\node[dot node, right of = b2',label=right:{$d$}] (q2') {};

\node[right of = c'] (lab_c) {(\romannumeral 4)};

\path[every node/.style={font=\sffamily\small}]

(a') edge[->] (b1')
(b1') edge[->] (b2')
edge[->] node[above] {\textcolor{red}{$q$}} (q1')
(b2') edge[->] (c')
edge[->] (q2')
(c') edge[->] (q2')
(q2') edge[->] node[right] {\textcolor{red}{$p'$}} (q1');

\coordinate[right of = q1',node distance=5mm] (st_arr);
\coordinate[right of = st_arr] (en_arr);
\path[every node/.style={font=\sffamily\small}]
(st_arr) edge[derives] node[below] {$t_4,t_5$}(en_arr);

\node[dot node, right of = en_arr,node distance=5mm,label=left:{$b$}] (b1) {};
\node[dot node, above of = b1,label=left:{$a$}] (a) {};
\node[dot node, below of = b1,label=left:{$b$}] (b2) {};
\node[dot node, below of = b2,label=left:{$c$}] (c) {};
\node[dot node, right of = b2,label=right:{$d$}] (d1) {};
\node[dot node, right of = b1,label=right:{$d$}] (d2) {};
\node[dot node, right of = a,label=right:{$e$}] (e) {};

\node[right of = c] (lab_d) {(\romannumeral 5)};

\path[every node/.style={font=\sffamily\small}]
(a) edge[->] node[left,blue] {$p$} (b1)
(b1) edge[->] node[left,blue] {$p$} (b2)
edge[->] node[above,blue] {$q$} (d2)
(b2) edge[->] node[left,blue] {$p$} (c)
edge[->] node[above,blue] {$q$} (d1)
(c) edge[->] node[below,blue] {$p'$} (d1)
(d1) edge[->] node[right,blue] {$p'$}(d2)
(d2) edge[->] node[right,blue] {$p'$} (e);

\node[dot node, right of = e,node distance=50mm,label=left:{$a$}] (a) {};
\node[dot node, below of = a,label=left:{$b$}] (b1) {};
\node[dot node, below of = b1,label=left:{$b$}] (b2) {};
\node[dot node, below of = b2,label=left:{$c$}] (c) {};
\node[dot node, right of = b2,label=right:{$d$}] (d1) {};
\node[dot node, right of = b1,label=right:{$d$}] (d2) {};
\node[dot node, right of = a,label=right:{$e$}] (e) {};

\node[right of = c,xshift=2mm] (lab_d) {(\romannumeral 7)};

\coordinate[left of = b1,xshift=2mm] (en_arr);
\coordinate[left of = en_arr,xshift=2mm] (st_arr);
\node[left of = st_arr,node distance=5mm] (q1') {};
\node[dot node, left of = q1',label=left:{$b$}] (b1') {};
\node[dot node, above of = b1',label=left:{$a$}] (a') {};
\node[dot node, below of = b1',label=left:{$b$}] (b2') {};
\node[dot node, below of = b2',label=left:{$c$}] (c') {};
\node[dot node, right of = b2',label=right:{$d$}] (q2') {};

\coordinate[left of = b1',xshift=2mm] (en_arr');
\coordinate[left of = en_arr',xshift=2mm] (st_arr');

\coordinate[left of=a',yshift=2mm,xshift=-8mm] (top_sep);
\coordinate[left of=c',yshift=-2mm,xshift=-8mm] (bottom_sep);

\node[right of = c'] (lab_c) {(\romannumeral 6)};

\path[every node/.style={font=\sffamily\small}]

(top_sep) edge[densely dotted,gray] (bottom_sep)

(a) edge[->] node[left,blue] {$p$} (b1)
(b1) edge[->] node[left,blue] {$p$} (b2)
edge[->] node[below,pos=0.6,blue] {$q$} (d1)
(b2) edge[->] node[left,blue] {$p$} (c)
edge[->] node[above,pos=0.6,blue] {$q$} (d2)
(c) edge[->] node[below,blue] {$p'$} (d1)
(d1) edge[->] node[right,blue] {$p'$} (d2)
(d2) edge[->] node[right,blue] {$p'$} (e)

(st_arr) edge[derives] node[below] {$t_4, t_5$} (en_arr)
(st_arr') edge[derives] node[below] {\begin{tabular}{c}$t_4$\\\textrm{from}\\\textrm{(\romannumeral 3)}\end{tabular}} (en_arr')

(a') edge[->] (b1')
(b1') edge[->] (b2')
edge[->] (q2')
(b2') edge[->] (c')
edge[->] node[pos=0.6,above,red] {$q$} (q1')
(c') edge[->] (q2')
(q2') edge[->] node[right,red] {$p'$} (q1');

\end{tikzpicture}
\caption{Two alternative derivations using the automaton of Example \ref{ex:automaton}. Parts (\romannumeral 1), (\romannumeral 2), and (\romannumeral 3) are common to both derivations; (\romannumeral 4) and (\romannumeral 5) represent one possible completion, while (\romannumeral 6) and (\romannumeral 7) represent an alternative completion. Light grey double edges denote derivation steps and are labeled with the corresponding transition(s). Red edge labels on partial graphs denote frontier states, while blue edge labels on complete graphs denote an accepting run.
\label{fig:aut_ex}}
\end{center}
\end{figure*}

For the remainder of this section and the next, we will focus only on non-planar DAG automata, and when we refer to DAG automata, we mean this type. To formally define them, we need a notation for multisets---sets that can contain repeated elements. A \textbf{multiset} is a pair $(S,m)$ where $S$ is a finite set and $m : S \rightarrow \mathbb{N}$ is a count function---that is, $m(x)$ counts the number of times $x$ appears in the multiset. The set of all finite multisets over $S$ is $M(S)$. When we write multisets, we will often simply enumerate their elements. For example, $\{p, q, q\}$ is the multiset containing one $p$ and two $q$'s, and since multisets are unordered, it can also be written $\{q, p, q\}$ or $\{q, q, p\}$. We write $\emptyset$ for a multiset containing no elements.

\begin{defn}
A \textbf{DAG automaton} is a triple $A = (Q,\Sigma,T)$ where $Q$ is a finite set of states; $\Sigma$ is a finite set of node labels; and $T$ is a finite set of transitions of the form $\alpha \xrightarrow{\sigma} \beta $ where $\sigma \in \Sigma$ is a node label, $\alpha\in M(Q)$ is the left-hand side, and $\beta \in M(Q)$ is the right-hand side.
\label{defn:daga}
\end{defn}

\begin{example}\label{ex:automaton}
Let $A = (Q,\Sigma,T)$ be a DAG automaton where $Q = \{p,p',q\}$, $\Sigma = \{a,b,c,d,e\}$ and 
the transitions in $T$ are as follows:

\noindent\begin{minipage}{.4\linewidth}
\begin{align*}
\emptyset &\xrightarrow{a} \{p\} \tag{$t_1$}\\
\{p\} &\xrightarrow{b} \{p,q\} \tag{$t_2$}\\
\{p\} &\xrightarrow{c} \{p'\} \tag{$t_3$}
\end{align*}
\end{minipage}\begin{minipage}{.15\linewidth}~\end{minipage}\begin{minipage}{.45\linewidth}
\begin{align*}
\{p',q\}&\xrightarrow{d} \{p'\} \tag{$t_4$}\\
\{p'\}&\xrightarrow{e} \emptyset  \tag{$t_5$}
\end{align*}
\end{minipage}
\end{example}

\subsubsection{Generating single-rooted DAGs\label{sec:gen_dag}}

A DAG automaton generates a graph from root to leaves. To illustrate this, we'll focus on the case where a DAG is allowed to have only a single root, and return to the multi-rooted case in Section \ref{sec:multi}. To generate the root, the DAG automaton can choose any transition with $\emptyset$ on its left-hand side---these transitions behave like transitions from the start state in a finite automaton on strings, and always generate roots. In our example, the only available transition is $t_1$, which generates a node labeled $a$ with a dangling outgoing edge in state $p$, as in Figure~\ref{fig:aut_ex}(\romannumeral 1). The set of all such dangling edges is the \textbf{frontier} of a partially-generated DAG. 

While there are edges on the frontier, the DAG automation must choose and apply a transition whose left-hand side matches some subset of them. In our example, the automaton can choose either $t_2$ or $t_3$, each matching the available $p$ edge. The edges associated with the matched states are attached to a new node with new outgoing frontier edges specified by the transition, and the matched states are removed from the frontier. If our automaton chooses $t_2$, it arrives at the configuration in Figure~\ref{fig:aut_ex}(\romannumeral 2), with a new node labeled $b$, new edges on the frontier labeled $p$ and $q$, and the incoming $p$ state forgotten. Once again, it must choose between $t_2$ and $t_3$---it cannot use the $q$ state because that state can only be used by $t_4$, which also requires a $p'$ on the frontier. So each time it applies $t_2$, the choice between $t_2$ and $t_3$ repeats.

If the automaton applies $t_2$ again and then $t_3$, as it has done in Figure~\ref{fig:aut_ex}(\romannumeral 3), it will face a new set of choices, between $t_4$ and $t_5$. But notice that choosing $t_5$ will leave the $q$ states stranded, leaving a partially derived DAG. We consider a run of the automaton successful only when the frontier is empty, so this choice leads to a dead end.

If the automaton chooses $t_4$, it has an additional choice: it can combine $p'$ with \emph{either} of the available $q$ states. If it combines with the lowermost $q$, it arrives at the graph in Figure~\ref{fig:aut_ex}(\romannumeral 4), and it can then apply $t_4$ to consume the remaining $q$, followed by $t_5$, which has $\emptyset$ on its right-hand side. Transitions to $\emptyset$ behave like transitions to a final state in a finite automaton, and generate leaf nodes, so we arrive at the complete graph in Figure \ref{fig:aut_ex}(\romannumeral 5). If  the $p'$ state in Figure~\ref{fig:aut_ex}(\romannumeral 3) had instead combined with the upper $q$, a different DAG would result, as in Figure \ref{fig:aut_ex}(\romannumeral 6-\romannumeral 7). 

\begin{center}
\begin{tikzpicture}[>=stealth',auto,node distance=9mm,
  square node/.style={draw,line width=0.25mm,rectangle,inner sep=2pt}
,dot node/.style={draw,circle,minimum size=1.5mm,fill,inner sep=0pt,outer sep=0pt},thick node/.style={draw,blue,thick,circle,minimum size=1.5mm,inner sep=0pt,outer sep=0pt},dashed node/.style={draw,ultra thick,densely dotted,red,circle,minimum size=1.5mm,inner sep=0pt,outer sep=0pt}]

\node[dot node,label=below:{$a$}] (a) {};
\node[dot node, right of = a,label=below:{$b$}] (b1) {};
\node[thick node, right of = b1,label=below:{$b$}] (b2) {};
\node[dashed node, right of = b2,label=below:{$b$}] (b3) {};
\node[thick node, right of = b3,label=below:{$b$}] (b4) {};
\node[dot node, right of = b4,label=below:{$c$}] (c) {};
\node[dashed node, above of = b4,label=above:{$d$}] (d1) {};
\node[thick node, left of = d1,label=above:{$d$}] (d2) {};
\node[dashed node, left of = d2,label=above:{$d$}] (d3) {};
\node[dot node, left of = d3,label=above:{$d$}] (d4) {};
\node[dot node, left of = d4,label=above:{$e$}] (e) {};

\path[every node/.style={font=\sffamily\small}]

(a) edge[->,densely dotted] (b1)
(b1) edge[->,densely dotted] (b2)
edge[->,densely dotted] (d4)
(b2) edge[->] (b3)
edge[->] (d1)
(b3) edge[->] (b4)
edge[->] (d2)
(b4) edge[->,densely dotted] (c)
edge[->] (d3)
(d4) edge[->,densely dotted] (e)
(c) edge[->,bend right] (d1)
(d1) edge[->] (d2)
(d2) edge[->] (d3)
(d3) edge[->] (d4); 
\end{tikzpicture}
\end{center}
The DAGs in Figure \ref{fig:aut_ex}(\romannumeral 5) and Figure \ref{fig:aut_ex}(\romannumeral 7) are planar, which means they can be drawn without crossing edges.\footnote{While the graph in Figure \ref{fig:aut_ex}(\romannumeral 7) is drawn with crossing $b-d$ edges, one of these edges can be redrawn so that they do not cross.} But this DAG automaton can also produce non-planar DAGs like the one above. To see that it is non-planar, we can contract the dotted edges and fuse their endpoints, giving us a minor subgraph called $K_{3,3}$, the complete (undirected) bipartite graph over two sets of three nodes---one set is denoted by hollow blue nodes, and the other by dotted red nodes. Any graph with a $K_{3,3}$ minor is non-planar \citep{wagner1937eigenschaft}.

\subsubsection{Recognising DAGs and DAG languages}
We define the language generated by a DAG automaton in terms of recognition, which asks if an input DAG could have been generated by an input automaton. We recognise a DAG by finding a run of the automaton that could have generated it. We guess a run on a DAG by guessing a state for each of its edges, and then ask whether those states simulate a valid sequence of transitions.

A \textbf{run} of a DAG automaton $A = (Q,\Sigma,T)$ on a DAG $G = (V,E,\text{lab},\text{src}, \text{tar})$ is a mapping $\rho: E \rightarrow Q$ from edges of $G$ to automaton states $Q$. We extend $\rho$ to multisets by saying $\rho(\{e_1,\dots,e_n\}) = \{\rho(e_1),\dots,\rho(e_n)\}$, and we call 
a run \textbf{accepting} if for all $v \in V$ there is a corresponding transition $\rho(\textsc{In}(v)) \xrightarrow{\text{lab}(v)}\rho(\textsc{Out}(v))$ in $T$. DAG $G$ is \textbf{recognised} by automaton $A$ if there is an accepting run of $A$ on $G$.

\begin{example}
The DAGs in Figure \ref{fig:aut_ex}(\romannumeral 5) and \ref{fig:aut_ex}(\romannumeral 7) are recognised by the automaton in Example \ref{ex:automaton}. The only accepting run for each DAG is denoted by the blue edge labels.
\end{example}

Given a DAG automaton $A$, its single-rooted language $L_s(A)$ is
$\{G \in \mathcal{G}_\Sigma^1\mid A \text{ recognizes }G\}$.

\subsection{Probability and weighted DAG automata}

\begin{defn}
Given a language $L$ of DAGs, a \textbf{probability distribution} over $L$ is any function $p: L \rightarrow \mathbb{R}$ meeting two requirements:
\begin{enumerate}
\item[\textbf{(R1)}] Every DAG must have a probability between 0 and 1, inclusive. Formally, we require  that for all $G \in L$, $p(G) \in [0,1]$.
\item[\textbf{(R2)}] The probabilities of all DAGs must sum to one. Formally, we require $\sum_{G \in L}p(G) = 1$.
\end{enumerate}
\end{defn}
R1 and R2 suffice to define a probability distribution, but in practice we need something stronger than R1: all DAGs must receive a \emph{non-zero} weight, since in practical applications, objects with probability zero are effectively not in the language.

\begin{defn}A probability distribution $p$ has \textbf{full support} of $L$ if and only if it meets condition R1'.
\begin{enumerate}
\item[\textbf{(R1')}] Every DAG must have a probability greater than 0 and less than or equal to 1. Formally, we require that for all $G \in L$, $p(G) \in (0,1]$.
\end{enumerate}
\end{defn}

While there are many ways to define a function that meets requirements R1' and R2, probability distributions in natural language processing are widely defined in terms of weighted automata or grammars, so we adapt a common definition of weighted grammars \citep{booth_thompson} to DAG automata.

\begin{defn}
A \textbf{weighted DAG automaton} is a pair $(A,w)$ where $A=(Q,\Sigma,T)$ is a DAG automaton and $w: T \rightarrow \mathbb{R}$ is a function that assigns real-valued weights to the transitions of $A$.
\end{defn}

Since weights are functions of transitions, we will write them on transitions following the node label and a slash ($/$). For example, if $p \xrightarrow{a} q$ is a transition and $2$ is its weight, we write $p \xrightarrow{a/2} q$.

\begin{example}\label{ex:weighted}
Let $(A,w)$ be a weighted DAG automaton with $A = (Q,\Sigma,T)$, where $Q = \{p,q\}$, $\Sigma = \{a,b,c\}$, and the weighted transitions of $T$ are as follows:

\noindent\begin{minipage}{.5\linewidth}
\begin{align*}
\emptyset &\xrightarrow{a/0.5} \{p,q\} \tag{$t'_1$}\\
\{p\} &\xrightarrow{b/0.5} \{p\}  \tag{$t'_2$} \\
\end{align*}
\end{minipage}\begin{minipage}{.05\linewidth}~\end{minipage}\begin{minipage}{.45\linewidth}
\begin{align*}
\{p,q\} &\xrightarrow{c/1} \emptyset  \tag{$t'_3$}\\
\end{align*}
\end{minipage}
\end{example}
\noindent We use the weights on transitions to weight runs.

\begin{defn} Given a weighted DAG automaton $(A,w)$ and a DAG $G = (V,E,\text{lab},\text{src}, \text{tar})$ with an accepting run $\rho$, we extend $w$ to compute the \textbf{weight of the run} $w(\rho)$ by multiplying the weights of all of its transitions:
\[w(\rho) = \prod_{v\in V}w(\rho(\textsc{In}(v))\xrightarrow{\text{lab}(v)}\rho(\textsc{Out}(v)))\]
\end{defn}

\begin{example}\label{ex:run_weight}

The DAG automaton of Example~\ref{ex:weighted} generates the DAG below, shown with its only accepting run in blue and the weighted transitions that generated it in grey. The weight of the accepting run is $0.5 \times 0.5 \times0.5 \times 1 = 0.125$.
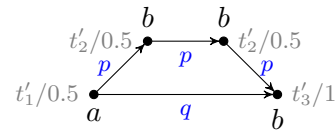
\begin{center}
\begin{tikzpicture}[>=stealth',auto,node distance=10mm,
  square node/.style={draw,line width=0.25mm,rectangle,inner sep=2pt}
,dot node/.style={draw,fill,circle,minimum size=1.1mm,inner sep=0pt,outer sep=0pt}]

\node[dot node, label=left:{\small \textcolor{gray}{$t_1'/0.5$}},label=below:{$a$}] (a) {};
\node[dot node, above right of = a, label=left:{\small \textcolor{gray}{$t_2'/0.5$}},label=above:{$b$}] (b1) {};
\node[dot node, right of = b1, label=right:{\small \textcolor{gray}{$t_2'/0.5$}},label=above:{$b$}] (b2) {};
\node[dot node, below right of = b2, label=right:{\small \textcolor{gray}{$t_3'/1$}},label=below:{$b$}] (c) {};

\path[every node/.style={font=\sffamily\small}]

(a) edge[->] node[left,blue] {$p$} (b1)
edge[->] node[below,blue] {$q$} (c)
(b1) edge[->] node[below,blue] {$p$} (b2)
(b2) edge[->] node[right,blue] {$p$} (c);
\end{tikzpicture}
\end{center}
\end{example}

Let $R_A(G)$ be the set of all accepting runs of a DAG $G$ using the automaton $A$. We extend $w$ to calculate the weight of a DAG $G$ as the sum of the weights of all the runs that produce it:
\[w(G) = \sum_{\rho\in R_A(G)}w(\rho).\]

While all weighted DAG automata assign real values to DAGs, not all weighted DAG automata define probability distributions. To do so, they must also satisfy requirements R1 and R2.

\begin{defn}
A weighted automaton $(A,w)$ over language $L(A)$ is \textbf{probabilistic} if and only if function $w: L(A) \rightarrow \mathbb{R}$ is a probability distribution.
\end{defn}
\begin{example}
Consider the weighted automaton in Example \ref{ex:weighted}. Every DAG generated by this automaton must use $t'_1$ and $t'_3$ exactly once, and can use $t'_2$ any number of times. If we let $G_n$ be the DAG that uses $t'_2$ exactly $n$ times, then the language $L$ defined by this automaton is $\bigcup_{n\in \mathbb{N}}G_n$. Since $w(G_n) = w(t'_1)w(t'_2)^nw(t'_3)$ and $w(t'_1)$, $w(t'_2)$ and $w(t'_3)$ are positive, $w$ satisfies R1 and:
\begin{align*}
\sum_{G \in L}w(G) &=  \sum_{n=0}^\infty w(G_n) = \sum_{n=0}^\infty w(t'_1)w(t'_2)^nw(t'_3)\\
&= \sum_{n=0}^\infty 0.5^{n+1} = 1
\end{align*}
Thus $w$ also satisfies R2 and the weighted automaton in Example \ref{ex:weighted} is probabilistic.
\end{example}
\begin{defn}
A probabilistic automaton $(A,w)$ over language $L(A)$ is \textbf{probabilistic with full support} if and only if $w$ has full support of $L(A)$.
\end{defn}

For every finite automaton over strings or trees, there is a weighting of its transitions that makes it probabilistic \citep{booth_thompson}, and it is easy to show that it can be made probabilistic with full support. For example, string finite automata have full support if for every state the sum of weights on its outgoing transitions is 1 and each weight is greater than 0.\footnote{Assuming no epsilon transitions, in our notation for DAG automata restricted to strings this would include transitions to $\emptyset$, which correspond to states with a final probability of 1 \citep{Mohri+etal:2008:hspsc}.} But as we will show, this is not always possible for DAG automata.

\section{Non-probabilistic DAG automata\label{sec:proof}}

We will exhibit a DAG automaton that generates factorially many DAGs for a given number of nodes, and we will show that for any nontrivial assignment of weights, this factorial growth rate causes the weight of all DAGs to sum to infinity.





\begin{thm}
Let $\mathcal{A}$ be the automaton defined in Example \ref{ex:automaton}. There is no $w$ that makes $(\mathcal{A}, w)$ probabilistic with full support over $L_s(\mathcal{A})$.
\begin{proof}
In any run of the automaton, transition $t_1$ is applied exactly once to generate the single root, placing a $p$ on the frontier. This gives a choice between $t_2$ and $t_3$. If the automaton chooses $t_2$, it keeps one $p$ on the frontier and adds a $q$, and must then repeat the same choice. Suppose it chooses $t_2$ exactly $n$ times in succession, and then chooses $t_3$. Then the frontier will contain $n$ edges in state $q$ and one in state $p'$.
The only way to consume all of the frontier states is to apply transition $t_4$ exactly $n$ times, consuming a $q$ at each step, and then apply $t_5$ to consume $p'$ and complete the derivation. Hence in any accepting run, $t_1, t_3$ and $t_5$ are each applied once, and $t_2$ and $t_4$ are each applied $n$ times, for some $n\geq 0$. Since transitions map uniquely to node labels, it follows that every DAG in $L_s(\mathcal{A})$ will have exactly one node each labeled $a$, $c$, and $e$; and $n$ nodes each labeled $b$ and $d$.

When the automaton applies $t_4$ for the first time, it has $n$ choices of $q$ states to consume, each distinguished by its unique path from the root. The second application of $t_4$ has $n-1$ choices of $q$, and the $i$th application of $t_4$ has $n-(i-1)$ choices. Therefore, there are $n!$ different ways to consume the $q$ states, each producing a unique DAG.


Let $f(n)$ be the weight of a run where $t_2$ has been applied $n$ times, and to simplify our notation, let $B =w(t_1)w(t_3)w(t_5)$, and $C = w(t_2)w(t_4)$.
Let $c(n)$ be the number of unique runs where $t_2$ has been applied $n$ times. By the above:
\begin{align*}
&f(n)=w(t_1)w(t_2)^nw(t_3)w(t_4)^nw(t_5) =BC^n\\
&c(n)=n!
\end{align*}

Now we claim that any DAG in $L_s(\mathcal{A})$ has exactly one accepting run, because the mapping of node labels to transitions also uniquely determines the state of each edge in an accepting run. For example, a $b$ node must result from a $t_2$ transition and a $d$ node from a $t_4$ transition, and since the output states of $t_2$ and input states of $t_4$ share only a $q$, any edge from a $b$ node to a $d$ node must be labeled $q$ in any accepting run. Now let $G\in L_s(\mathcal{A})$ be a DAG with $n$ nodes labeled $b$. Since $G$ has only one accepting run, we have:
\[w(G) = f(n)\]

Let $L_n$ be the set of all DAGs in $L_s(\mathcal{A})$ with $n$ nodes labeled $b$. Then $L_s(\mathcal{A}) = \bigcup_{n =0}^\infty L_n$ and:
\begin{align*}
\sum_{G\in L_s(\mathcal{A})}w(G) = \sum_{n = 0}^\infty \sum_{G\in L_n} w(G) &= \sum_{n=0}^\infty  c(n)f(n) \\
&= \sum_{n=0}^\infty (n!)\big(BC^n\big)
\end{align*}

Hence for $(\mathcal{A},w)$ to be probabilistic with full support, R1' and R2 require us to choose $B$ and $C$ so that $BC^n \in (0,1]$ and $\sum_{n=0}^\infty n!BC^n = 1$ for all $n$. Note that this does not constrain the component weights of $B$ or $C$ to be in $(0,1]$---they can be any real numbers. But since we require $BC^n$ to be positive for all $n$, both $B$ and $C$ must also be positive. If either were 0, then $BC^n$ would be 0 for $n>0$; if either were negative, then $BC^n$ would be negative for some or all values of $n$. 

Now we show that any choice of positive $C$ causes $\sum_{G\in L_s(\mathcal{A})}w(G)$ to diverge.
Given an infinite series of the form $\sum_{n=0}^\infty a_n$, the \textbf{ratio test} \citep{ratio} considers the ratio between adjacent terms in the limit, $\lim_{n\rightarrow\infty}\frac{|a_{n+1}|}{|a_n|}$. If this ratio is greater than 1, the series diverges; if less than 1 the series converges; if exactly 1 the test is inconclusive. In our case:
\[\lim_{n\rightarrow \infty}\frac{|(n+1)!BC^{n+1}|}{|n!BC^n|} = \lim_{n\rightarrow \infty}(n+1)|C|=\infty.\]
Hence $\sum_{G\in L_s(\mathcal{A})}$ diverges for any choice of $C$, equivalently for any choice of weights. So there is no $w$ for which $(\mathcal{A},w)$ is probabilistic with full support over $L_s(\mathcal{A})$.
\end{proof}
\label{propn:single}
\end{thm}

Note that any automaton recognising $L_s(\mathcal{A})$ must accept factorially many DAGs in the number of nodes. Our proof implies that there is \emph{no} probabilistic DAG automaton for language $L_s(\mathcal{A})$, since no matter how we design its transitions---each of which must be isomorphic to one in $\mathcal{A}$ apart from the identities of the states---the factorial will eventually overwhelm the constant factor corresponding to $C$ in our proof, no matter how small it is.

Theorem~\ref{propn:single} does not rule out \emph{all} probabilistic variants of $\mathcal{A}$. It requires R1'---if we only require the weaker R1, then a solution of B=1 and C=0 makes the automaton probabilistic. But this trivial distribution is not very useful: it assigns all of its mass to the singleton language \{\tikz[baseline=0.5ex]{
  \node[circle,fill,minimum size=1.1mm,inner sep=0pt,outer sep=0pt,draw,label=above:{\small a}] (a) {};
  \node[circle,fill,minimum size=1.1mm,inner sep=0pt,outer sep=0pt,draw,right of=a,label=above:{\small c}] (c) {};
  \node[circle,fill,minimum size=1.1mm,inner sep=0pt,outer sep=0pt,draw,right of=c,label=above:{\small e}] (e) {};
  \draw[->] (a) -- (c);
  \draw[->] (c) -- (e);
}\}.

Theorem~\ref{propn:single} also does not mean that it is impossible to define a probability distribution over $L_s(\mathcal{A})$ with full support. If, for every DAG $G$ with $n$ nodes labeled $b$, we let $p(G) = \frac{1}{2^{n+1}n!}$, then:
\[\sum_{G\in L_s(\mathcal{A})}w(G) = \sum_{n=0}^\infty \frac{1}{2^{n+1}n!}n! = \sum_{n=0}^\infty \frac{1}{2^{n+1}} = 1\] 
But this distribution does not factor over transitions, so it cannot be used with the dynamic programming algorithms of \citet{CL}.

A natural way to define distributions using a DAG automaton is to define two conditional probabilities: one over the choice of nodes to rewrite, given a frontier; and one over the choice of transition, given the chosen nodes. The latter factors over transitions, but the former does not, so it also cannot use the algorithms of \citet{CL}.\footnote{In this model, the subproblems of a natural dynamic program depend on the set of possible frontiers, rather than subsets of nodes as in the algorithms of \citet{CL}. We do not know whether this could be made efficient.}

Theorem \ref{propn:single} only applies to single-rooted, non-planar DAG automata of bounded degree. Next we ask whether it extends to other DAG automata, including those that recognise multi-rooted DAGs, DAGs of unbounded degree, and planar DAGs.

\subsection{Multi-rooted DAGs\label{sec:multi}}
What happens when we consider DAG languages that allow multiple roots? In one reasonable interpretation of AMRbank, over three quarters of the DAGs have multiple roots \citep{Kuhlmann_graphbank}, so we want a model that permits this.\footnote{AMR annotations are single-rooted, but they achieve this by duplicating edges: every edge type, like \textsc{Arg0}, has an inverse type, like \textsc{Arg0-Of}. The number cited here assumes edges of the second type are converted to the first type by reversing their direction.}

Section \ref{sec:gen_dag} explained how a DAG automaton can be constrained to generate single-rooted languages, by restricting start transitions (i.e. those with $\emptyset$ on the left-hand side) to a single use at the start of a derivation. To generate DAGs with multiple roots, we simply allow start transitions to be applied at any time. We still require the resulting DAGs to be connected. For an automaton $A$, we define its multi-rooted language $L_m(A)$ as $\{G \in \mathcal{G}_\Sigma^* | A \text{ recognises } G\}$.

Although one automaton can define both single- and multi-rooted DAG languages, these languages are incomparable. \citet{drewes_mol} uses a construction very similar to the one in Theorem~\ref{propn:single} to show that single-rooted languages have very expressive path languages, which he argues are too expressive for modeling semantics.\footnote{The \textbf{path language} of a DAG is the set of strings that label a path from a root to a leaf, and the path language of a DAG language is the set of all such strings over all DAGs. For example, the path language of the DAG in Figure~\ref{fig:aut_ex}(\romannumeral 5) is $\{abde, abbdde, abbcdde\}$.  \citet{drewes_parikh} show that path languages of multi-rooted DAG automata are regular, while those of single-rooted DAG automata characterised by a partially blind multi-counter automaton.} Since the constructions are so similar, it natural to wonder if the problem that single-rooted automata have with probabilities is related to their problem with expressivity, and whether it likewise disappears when we allow multiple roots. We now show that multi-rooted languages have the same problem with probability, because any multi-rooted language contains the single-rooted language as a sublanguage. 

\begin{corollary}\label{propn:multi}
Let $\mathcal{A}$ be the automaton defined in Example \ref{ex:automaton}. There is no $w$ that makes $(\mathcal{A}, w)$ probabilistic with full support over $L_m(\mathcal{A})$.
\begin{proof}

By their definitions, $L_s(\mathcal{A}) \subset L_m(\mathcal{A})$, so:
\begin{multline*}
\sum_{G\in L_m(\mathcal{A})}w(G) =\hfill \\\hfill \sum_{G\in L_s(\mathcal{A})}w(G) + \sum_{G\in L_m(\mathcal{A}) \setminus L_s(\mathcal{A})}w(G)
\end{multline*} 
The first term is $\infty$ by Theorem \ref{propn:single} and the second is positive by R1', so the sum diverges, and there is no $w$ for which $(\mathcal{A},w)$ is probabilistic with full support over $L_m(\mathcal{A})$.
\end{proof}
\end{corollary}

\subsection{DAGs of unbounded degree\label{sec:unbounded}}
The maximum degree of any node in any DAG recognised by a DAG automaton is bounded by the maximum number of states in any transition, because any transition $\alpha \xrightarrow{\sigma} \beta$ generates a node with $|\alpha|$ incoming edges and $|\beta|$ outgoing edges. So, the families of DAG languages we have considered all have bounded degree. 

DAG languages with unbounded degree could be useful to model phenomena like coreference in meaning representations, and they have been studied by \citet{Quernheim:2012:TPA:2392936.2392948} and \citet{CL}. These families generalise and strictly contain the family of bounded-degree DAG languages, so they too, include DAG automata that cannot be made probabilistic.

\subsection{Implications for semantic DAGs}
We introduced DAG automata as a tool for modeling the meaning of natural language, but the DAG automaton in Theorem~\ref{propn:single} is very artificial, so it's natural to ask whether it has any real relevance to natural language. We will argue informally that this example illustrates a pervasive problem with DAG automata---specifically, we conjecture that the factorial growth we observe in Theorem~\ref{propn:single} arises under very mild conditions that arise naturally in models of AMR.

Consider a model of object control in a sentence like ``I help Ruby help you'' and its AMR below. 
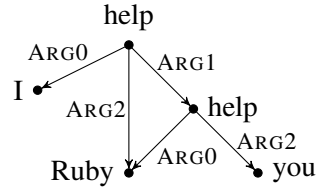
\begin{center}
\begin{tikzpicture}[>=stealth',auto,node distance=12mm,
  square node/.style={draw,line width=0.25mm,rectangle,inner sep=2pt}
,dot node/.style={draw,circle,fill,minimum size=1.1mm,inner sep=0pt,outer sep=0pt}]
\node[dot node,label=above:{help}] (h) {};
\node[dot node, left of = h, yshift=-6mm, label=left:{I}] (i) {};
\node[dot node, below right of = h,label=right:{help}] (h') {};
\node[dot node, below left of = h',label=left:{Ruby}] (y) {};
\node[dot node, below right of = h',label=right:{you}] (ruby) {};

\path[every node/.style={font=\small}]

(h) edge[->] node[right,xshift=-0.2cm,yshift=0.2cm] {\textsc{Arg1}} (h')
edge[->] node[left,xshift=0.2cm,yshift=0.2cm] {\textsc{Arg0}} (i)
edge[->] node[left,xshift=0.1cm] {\textsc{Arg2}} (y)
(h') edge[->] node[right,xshift=-0.2cm,yshift=-0.2cm] {\textsc{Arg0}} (y)
edge[->] node[right] {\textsc{Arg2}} (ruby);
\end{tikzpicture}
\end{center}

We can extend the control structure unboundedly with additional helpers, as in ``I help Briony help Kim-Joy help Ruby help you'', and this leads to unboundedly long repetitive graphs like the one below. These graphs can be cut to separate the sequence of ``help'' predicates from their arguments, as illustrated by the dashed blue line.
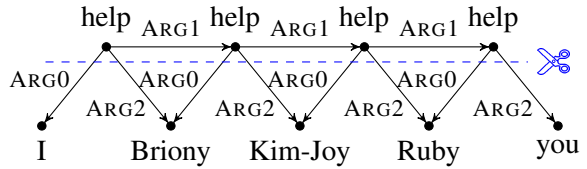
\begin{center}
\begin{tikzpicture}[>=stealth',auto,node distance=12mm,
dot node/.style={draw,circle,fill,minimum size=1.1mm,inner sep=0pt,outer sep=0pt}]
\node[dot node, label=below:{I}] (i) {};
\node[dot node, label=above:{help}, above right of=i,yshift=2mm] (help1) {};
\node[dot node, label=below:{Briony}, below right of=help1,yshift=-2mm] (briony) {};
\node[dot node, label=above:{help}, above right of=briony,yshift=2mm] (help2) {};
\node[dot node, label=below:{Kim-Joy}, below right of=help2,yshift=-2mm] (kimjoy) {};
\node[dot node, label=above:{help}, above right of=kimjoy,yshift=2mm] (help3) {};
\node[dot node, label=below:{Ruby}, below right of=help3,yshift=-2mm] (ruby) {};
\node[dot node, label=above:{help}, above right of=ruby,yshift=2mm] (help4) {};
\node[dot node, label=below:{you}, below right of=help4,yshift=-2mm] (you) {};

\coordinate[xshift=-8mm,yshift=-2mm] (cut_st) at (help1);
\node[xshift=8mm,yshift=-2mm,blue] (cut_en) at (help4) {\ScissorHollowLeft};

\path[every node/.style={font=\sc\small}]
(help1) edge[->] node[above] {Arg1} (help2)
edge[->] node[left,pos=0.4] {Arg0} (i)
edge[->] node[left,pos=0.8] {Arg2} (briony)
(help2) edge[->] node[above] {Arg1} (help3)
edge[->] node[left,pos=0.4] {Arg0} (briony)
edge[->] node[left,pos=0.8] {Arg2} (kimjoy)
(help3) edge[->] node[above] {Arg1} (help4)
edge[->] node[left,pos=0.4] {Arg0} (kimjoy)
edge[->] node[left,pos=0.8] {Arg2} (ruby)
(help4) 
edge[->] node[left,pos=0.4] {Arg0} (ruby)
edge[->] node[left,pos=0.8] {Arg2} (you)
(cut_st) edge[dashed,blue] (cut_en);
;
\end{tikzpicture}
\end{center}
Let a \textbf{cut} be a set of edges such that removing them splits the graph into two connected subgraphs: one containing the root, and the other containing all the leaves. Any cut in a complete graph could have been the frontier of a partially-derived graph. What if the number of edges in a cut---or \textbf{cut-width}---can be unbounded, as in this example? 

Since a DAG automaton can have only a finite number of states, there is some state that can occur unboundedly many times in a graph cut. All edges in a cut with this state can be rewired by permuting their target nodes, and the resulting graph will still be recognised by the automaton, since the rewiring would not change the multiset of states into or out of any node. If each possible rewiring results in a unique graph, then the number of recognised graphs will be factorial in the number of source nodes for these edges, and the argument of Theorem~\ref{propn:single} can be generalised to show that no weighting of any DAG automaton over the graph language makes it probabilistic with full support. For example, in the graph above, all possible rewirings of the \textsc{Arg2} edges result in a unique graph---and most do not model object control. Although edge labels are not states, their translation into node labels implies that they can only be associated to a finite number of transitions, hence to states. A full investigation of  conditions under which Theorem~\ref{propn:single} generalises is beyond the scope of this paper.

\begin{conjecture}
Under mild conditions, if language $L(A)$ of a DAG automaton $A$ has unbounded cut-width, there is no $w$ that makes $(A,w)$ probabilistic with full support.
\end{conjecture}

\section{Planar DAG automata\label{sec:planar}}
The fundamental problem with trying to assign probabilities to non-planar DAG automata is the factorial growth in the number of DAGs with respect to the number of nodes. Does this problem occur in planar DAG automata?

Planar DAG automata are similar to the DAG automata of Section \ref{sec:prelim} but with an important difference: they transition between \emph{ordered} sequences of states rather than unordered multisets of states. We write these sequences in parentheses, and their order matters: $(p,q)$ differs from $(q,p)$. We write $\epsilon$ for the empty sequence. When a planar DAG automaton generates DAGs, it keeps a strict order over the set of frontier states at all times. A transition whose left-hand side is $(p,q)$ can only be applied to adjacent states $p$ and $q$ in the frontier, with $p$ preceding $q$. The matched states are replaced in the frontier by the sequence of states in the transition's right-hand side, maintaining order.

\begin{example}\label{ex:planar_automaton}
Consider a planar DAG automaton with the following transitions:\\
\begin{minipage}{.4\linewidth}
\begin{align*}
\epsilon &\xrightarrow{a} (p) \tag{$t_1'''$} \\
(p) &\xrightarrow{b} (p,q) \tag{$t_2'''$} \\
(p) &\xrightarrow{c} (p') \tag{$t_3'''$} 
\end{align*}
\end{minipage}\begin{minipage}{.15\linewidth}~\end{minipage}\begin{minipage}{.45\linewidth}
\begin{align*}
(p',q) &\xrightarrow{d} (p') \tag{$t_4'''$} \\
(p') &\xrightarrow{e} \epsilon \tag{$t_5'''$} 
\end{align*}
\end{minipage}
\end{example}

In the non-planar case, $n$ applications of $t_2$ can generate $n!$ unique DAGs, but $n$ applications of the corresponding transition $t_2'''$ in this automaton can only generate one DAG. To see this, consider the partially derived DAG on the left below, with its frontier drawn in order from left to right. The $p'$ state can only combine with the $q$ state immediately to its right, and since dead-ends are not allowed, the only possible choice is to apply $t_4'''$ twice followed by $t_5'''$, so the DAG on the right is the only possible completion of the derivation.
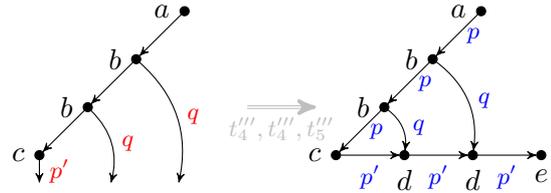
\begin{center}
\begin{tikzpicture}[>=stealth',auto,node distance=0.9cm,
  square node/.style={draw,line width=0.25mm,rectangle,inner sep=2pt}
,dot node/.style={draw,circle,fill,minimum size=1.2mm,inner sep=0pt,outer sep=0pt}]

\node[dot node,label=left:{$a$}] (a) {};
\node[dot node, below left of = a,label=left:{$b$}] (b1) {};
\node[dot node, below left of = b1,label=left:{$b$}] (b2) {};
\node[dot node, below left of = b2,label=left:{$c$}] (c) {};
\node[dot node, right of = c,label=below:{$d$}] (d1) {};
\node[dot node, right of = d1,label=below:{$d$}] (d2) {};
\node[dot node, right of = d2,label=below:{$e$}] (e) {};

\coordinate[left of = b2] (en_arr);
\coordinate[left of = en_arr] (st_arr);
\node[dot node, left of = st_arr, xshift=-1.2cm,label=left:{$b$}] (b2') {};
\node[dot node, above right of = b2',label=left:{$b$}] (b1') {};
\node[dot node, above right of = b1',label=left:{$a$}] (a') {};
\node[dot node, below left of = b2',label=left:{$c$}] (c') {};
\node[below of = c',node distance=5mm] (p') {};
\node[right of = p'] (q2') {};
\node[right of = q2'] (q1') {};

\path[every node/.style={font=\sffamily\small}]

(a) edge[->] node[right,blue] {$p$} (b1)
(b1) edge[->] node[right,blue] {$p$} (b2)
(b2) edge[->] node[right,blue] {$p$} (c)
(c) edge[->] node[below,blue] {$p'$} (d1)
(d1) edge[->] node[below,blue] {$p'$} (d2)
(d2) edge[->] node[below,blue] {$p'$} (e)
(b1) edge[->, bend left] node[right,blue] {$q$} (d2)
(b2) edge[->, bend left] node[right,blue] {$q$} (d1)

(st_arr) edge[->,double,lightgray] node[below] {$t_4''',t_4''',t_5'''$} (en_arr)

(a') edge[->] node[left] {} (b1')
(b1') edge[->] node[left] {} (b2')
(b2') edge[->] node[left] {} (c')
(c') edge[->] node[right,red] {$p'$} (p')
(b1') edge[->, bend left] node[right,red] {$q$} (q1')
(b2') edge[->, bend left] node[right,red] {$q$} (q2');
\end{tikzpicture}
\end{center}

This automaton is probabilistic when $w(t_1''') = w(t_2''') = \nicefrac{1}{2}$, $w(t_3''') = w(t_4''') = w(t_5''') = 1$, and indeed the argument in Theorem \ref{propn:single} does not apply to planar automata since the number of applicable transitions is linear in the size of the frontier. But planar DAG automata have other problems that make them unsuitable for modeling AMR.

The first problem is that there are natural language constructions that naturally produce non-planar DAGs in AMR. For example, consider the sentence ``Four contestants mixed, baked and ate cake.'' Its AMR, shown below, is not planar because it has a minor isomorphic to $K_{3,3}$, denoted by blue hollow and red dotted nodes. Any coordination of three predicates that share two arguments produces this structure. In the first release of AMR, 117 out of 12844 DAGs are non-planar.

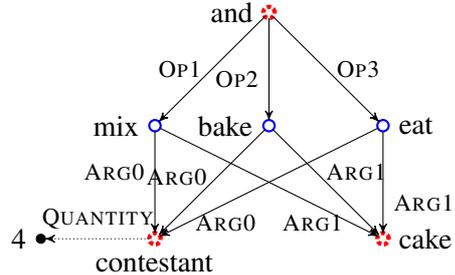
\begin{figure}[ht]
\begin{center}
\begin{tikzpicture}[>=stealth',auto,node distance=1.5cm,
  square node/.style={draw,line width=0.25mm,rectangle,inner sep=2pt}
,dot node/.style={draw,circle,fill,minimum size=1.2mm,inner sep=0pt,outer sep=0pt}
,thick node/.style={draw,blue,thick,circle,minimum size=1.5mm,inner sep=0pt,outer sep=0pt}
,dashed node/.style={draw,ultra thick,densely dotted,red,circle,minimum size=1.5mm,inner sep=0pt,outer sep=0pt}]
\node[dashed node, label=left:{and}] (and) {};
\node[thick node, below of = and, label=left:{bake}] (kick) {};
\node[thick node, left of = kick, label=left:{mix}] (run) {};
\node[thick node, right of = kick, label=right:{eat}] (beat) {};
\node[dashed node, below of = run, label=below:{contestant}] (person) {};
\node[dot node, left of = person,label=left:{4}] (3) {};
\node[dashed node, below of = beat, label=right:{cake}] (he) {};

\path[every node/.style={font=\small}]
(and) edge[->] node[left] {\sc Op1} (run)
edge[->] node[left,pos=0.6] {\sc Op2} (kick)
edge[->] node[right] {\sc Op3} (beat)
(run) edge[->] node[pos=0.4,left] {\sc Arg0} (person)
edge[->] node[below,pos=0.7] {\sc Arg1} (he)
(kick) edge[->] node[pos=0.45,left] {\sc Arg0} (person)
edge[->] node[right,pos=0.4] {\sc Arg1} (he)
(beat) edge[->] node[pos=0.7,below] {\sc Arg0} (person)
edge[->] node[right,pos=0.7] {\sc Arg1} (he)
(person) edge[densely dotted,->] node[above] {\sc Quantity} (3);
\end{tikzpicture}
\end{center}
\end{figure}

The second problem is that planar DAG automata model Type-0 string derivations by design \cite{kamimura}. This seems more expressive than needed to model natural language and means that many important decision problems are undecidable---for example, emptiness, which is decidable in polynomial time for non-planar DAG automata \citep{CL}.

\section{Conclusions}
The table below summarises the properties of several different variants of DAG automata. It has been argued that all of these properties are desirable for models of meaning representations \citep{drewes_mol}, so the table suggests that none of these formalisms are good candidates. We believe other formalisms may be more suitable, including several subfamilies of hyperedge replacement grammars \citep{drewes} that have recently been proposed \citep{drewes_rock,Matheja2015,mol_17}. 
\begin{center}
\begin{tabular}{c|c|c|c|c|c}\toprule
& \multicolumn{4}{c|}{non-planar} & planar\\ \cline{2-6}
bounded degree & \multicolumn{2}{c|}{yes} & \multicolumn{2}{c|}{no} & yes\\ \cline{2-6} roots & 1 & 1+ & 1 & 1+ & 1\\ \cline{1-6}
probabilistic & no & no & no & no & ? \\  
decidable & yes & yes & yes & yes & no \\  
regular paths & no & yes & no & yes & no \\  \bottomrule
\end{tabular}
\end{center}

\section*{Acknowledgements}

This work was supported in part by the EPSRC Centre for Doctoral Training in Data Science, funded by the UK Engineering and Physical Sciences Research Council (grant EP/L016427/1) and the University of Edinburgh. We thank Esma Balkir, Nikolay Bogoychev, Shay Cohen, Marco Damonte, Federico Fancellu, Joana Ribeiro, Nathan Schneider, Milo\v{s} Stanojevi\'{c}, Ida Szubert, Clara Vania, and the anonymous reviewers for helpful discussion of this work and comments on previous drafts of the paper.

\bibliography{biblio}

\begin{thebibliography}{20}
\expandafter\ifx\csname natexlab\endcsname\relax\def\natexlab#1{#1}\fi

\bibitem[{Banarescu et~al.(2013)Banarescu, Bonial, Cai, Georgescu, Griffitt,
  Hermjakob, Knight, Koehn, Palmer, and Schneider}]{amr}
Laura Banarescu, Claire Bonial, Shu Cai, Madalina Georgescu, Kira Griffitt, Ulf
  Hermjakob, Kevin Knight, Philipp Koehn, Martha Palmer, and Nathan Schneider.
  2013.
\newblock Abstract meaning representation for sembanking.
\newblock In \emph{Proceedings of the 7th Linguistic Annotation Workshop and
  Interoperability with Discourse}, pages 178--186, Sofia, Bulgaria.

\bibitem[{Berglund et~al.(2017)Berglund, Bj\"{o}rklund, and
  Drewes}]{drewes_parikh}
Martin Berglund, Henrik Bj\"{o}rklund, and Frank Drewes. 2017.
\newblock Single-rooted dags in regular dag languages: Parikh image and path
  languages.
\newblock In \emph{Proceedings of the 13th International Workshop on Tree
  Adjoining Grammars and Related Formalisms}, pages 94--101, Ume\r{a}, Sweden.

\bibitem[{Bj{\"{o}}rklund et~al.(2016)Bj{\"{o}}rklund, Drewes, and
  Ericson}]{drewes_rock}
Henrik Bj{\"{o}}rklund, Frank Drewes, and Petter Ericson. 2016.
\newblock Between a rock and a hard place - uniform parsing for hyperedge
  replacement {DAG} grammars.
\newblock In \emph{Language and Automata Theory and Applications - 10th
  International Conference, {LATA} 2016, Prague, Czech Republic, March 14-18,
  2016, Proceedings}, pages 521--532.

\bibitem[{Blum and Drewes(2016)}]{DBLP:conf/lata/BlumD16}
Johannes Blum and Frank Drewes. 2016.
\newblock Properties of regular {DAG} languages.
\newblock In \emph{Language and Automata Theory and Applications - 10th
  International Conference, {LATA} 2016, Prague, Czech Republic, March 14-18,
  2016, Proceedings}, pages 427--438.

\bibitem[{Booth and Thompson(1973)}]{booth_thompson}
T.L. Booth and R.A. Thompson. 1973.
\newblock Applying probability measures to abstract languages.
\newblock \emph{IEEE Transactions on Computers}, 22(5):442--450.

\bibitem[{Chiang et~al.(2013)Chiang, Andreas, Bauer, Hermann, Jones, and
  Knight}]{chiang}
David Chiang, Jacob Andreas, Daniel Bauer, Karl~Moritz Hermann, Bevan Jones,
  and Kevin Knight. 2013.
\newblock Parsing graphs with hyperedge replacement grammars.
\newblock In \emph{Proceedings of the 51st Annual Meeting of the Association
  for Computational Linguistics (Volume 1: Long Papers)}, pages 924--932,
  Sofia, Bulgaria.

\bibitem[{Chiang et~al.(2018)Chiang, Drewes, Gildea, Lopez, and Satta}]{CL}
David Chiang, Frank Drewes, Daniel Gildea, Adam Lopez, and Giorgio Satta. 2018.
\newblock Weighted {DAG} automata for semantic graphs.
\newblock \emph{Computational linguistics}, 44(1).

\bibitem[{D'{A}lembert(1768)}]{ratio}
Jean D'{A}lembert. 1768.
\newblock \emph{Opuscules}, volume~V.

\bibitem[{Drewes(2017)}]{drewes_mol}
Frank Drewes. 2017.
\newblock Dag automata for meaning representation.
\newblock In \emph{Proceedings of the 15th Meeting on the Mathematics of
  Language}, pages 88--99, London, UK.

\bibitem[{Drewes et~al.(1997)Drewes, Kreowski, and Habel}]{drewes}
Frank Drewes, Hans-J{\"o}rg Kreowski, and Annegret Habel. 1997.
\newblock Hyperedge replacement graph grammars.
\newblock In Grzegorz Rozenberg, editor, \emph{Handbook of Graph Grammars and
  Computing by Graph Transformation}, pages 95--162. World Scientific.

\bibitem[{Flanigan et~al.(2016)Flanigan, Dyer, Smith, and
  Carbonell}]{Flanigan+etal:2016:naacl}
Jeffrey Flanigan, Chris Dyer, Noah~A. Smith, and Jaime Carbonell. 2016.
\newblock Generation from abstract meaning representation using tree
  transducers.
\newblock In \emph{Proc. of NAACL-HLT}, pages 731--739.

\bibitem[{Gilroy et~al.(2017)Gilroy, Lopez, Maneth, and Simonaitis}]{mol_17}
Sorcha Gilroy, Adam Lopez, Sebastian Maneth, and Pijus Simonaitis. 2017.
\newblock ({R}e)introducing regular graph languages.
\newblock In \emph{Proceedings of the 15th Meeting on the Mathematics of
  Language (MoL 15)}, pages 100--113.

\bibitem[{Kamimura and Slutzki(1981)}]{kamimura}
Tsutomu Kamimura and Giora Slutzki. 1981.
\newblock Parallel and two-way automata on directed ordered acyclic graphs.
\newblock \emph{Information and Control}, 49(1):10--51.

\bibitem[{Kuhlmann and Oepen(2016)}]{Kuhlmann_graphbank}
Marco Kuhlmann and Stephan Oepen. 2016.
\newblock Squibs: Towards a catalogue of linguistic graph banks.
\newblock \emph{Computational Linguistics}, 42(4):819--827.

\bibitem[{Matheja et~al.(2015)Matheja, Jansen, and Noll}]{Matheja2015}
Christoph Matheja, Christina Jansen, and Thomas Noll. 2015.
\newblock \emph{Tree-Like Grammars and Separation Logic}, pages 90--108.
  Springer International Publishing, Cham.

\bibitem[{May et~al.(2010)May, Knight, and
  Vogler}]{May:2010:EIT:1858681.1858789}
Jonathan May, Kevin Knight, and Heiko Vogler. 2010.
\newblock Efficient inference through cascades of weighted tree transducers.
\newblock In \emph{Proceedings of the 48th Annual Meeting of the Association
  for Computational Linguistics}, ACL '10, pages 1058--1066.

\bibitem[{Mohri et~al.(2008)Mohri, Pereira, and Riley}]{Mohri+etal:2008:hspsc}
Mehryar Mohri, Fernando C.~N. Pereira, and Michael Riley. 2008.
\newblock Speech recognition with weighted finite-state transducers.
\newblock In Larry Rabiner and Fred Juang, editors, \emph{Handbook on Speech
  Processing and Speech Communication, Part {E}: Speech recognition}, pages
  69--88. Springer.

\bibitem[{van Noord and Bos(2017)}]{vanNoord}
Rik van Noord and Johan Bos. 2017.
\newblock Neural semantic parsing by character-based translation: Experiments
  with abstract meaning representations.
\newblock \emph{Computational Linguistics in the Netherlands Journal},
  7:93--108.

\bibitem[{Quernheim and Knight(2012)}]{Quernheim:2012:TPA:2392936.2392948}
Daniel Quernheim and Kevin Knight. 2012.
\newblock Towards probabilistic acceptors and transducers for feature
  structures.
\newblock In \emph{Proceedings of the Sixth Workshop on Syntax, Semantics and
  Structure in Statistical Translation}, SSST-6 '12, pages 76--85.

\bibitem[{Wagner(1937)}]{wagner1937eigenschaft}
Klaus Wagner. 1937.
\newblock {\"U}ber eine eigenschaft der ebenen komplexe.
\newblock \emph{Mathematische Annalen}, 114(1):570--590.

\end{thebibliography}
\bibliographystyle{acl_natbib}

\end{document}